# Low temperature Synchrotron X-ray diffraction studies on spin lattice coupling in $Co_3TeO_6$ and $Co_{2.5}Mn_{0.5}TeO_6$

Harishchandra Singh,[a,b,♣] A. K. Sinha,[a,b,♣] M. N. Singh,[a] A. Upadhyay,[a] Archna Sagdeo[a,b]

[a]*Indus Synchrotrons Utilization Division, Raja Ramanna Centre for Advanced Technology, Indore – 452013, India*

[b]*Homi Bhabha National Institute, Anushaktinagar, Mumbai-400094, India.*

In the quest of understanding significant variations in the physical, chemical and electronic properties of the novel functional materials, low temperature Synchrotron X-ray Diffraction (LT-SXRD) measurements on $Co_3TeO_6$ (CTO; a type-II) and $Co_{2.5}Mn_{0.5}TeO_6$ (CMTO; a type-I) multiferroics are presented. Magnetic phase diagram of CTO shows multiple magnetic transitions at zero fields, whereas, in CMTO, 20 K enhancement in the antiferromagnetic transition temperature is observed followed by near room temperature Griffiths phase. Rietveld analysis on LT-SXRD data of both the samples indicates important observations. For both CTO and CMTO, the magnetic anomalies are followed by structural anomalies, which is a clear signatures of spin lattice coupling and it's positive shift from CTO (26 K) to CMTO (45 K).

1. **Introduction**

Investigating novel materials at variable temperatures opens new frontiers to chemistry and physics of a substance, leading to the various exciting phenomena including phase transitions (Chung et al., 2000). Except a few (Scott, 2013), such materials exhibit functionality related to other magnetic, electric, dielectric and electronic properties mostly at low temperatures (Chung et al., 2000; Darligton and Megaw, 1973; Dernie and Marezio, 1970; Ren et al., 1998; Li et al., 2007; Schiffer et al., 1995). Of them, the magnetic behaviour such as magnetism, superconductivity etc. of the material along with their structrural correlation has drawn much attention during the last few decades (Chung et al., 2000 - Lee et al., 1995). There has also been a debate for several existing materials regarding probing their low temperature weak magnetic anomalies by temperature dependent structrural techniques. In order to attempt the same, we present

[♣]Corresponding author. Tel.: +91-731-244 2583/2125/2155; Fax: +91-731-244 2140.
E-mail address: singh85harish@gmail.com, anil@rrcat.gov.in (for beamline correspondence)

low temperature synchrotron X-ray diffraction (LT-SXRD) and DC magnetization studies on two multiferroic materials i.e. $Co_3TeO_6$ (CTO; a type-II) (Becker et al., 2006; Golubko et al., 2010; Ivanov et al., 2012; Hudl et al., 2011; Her et al., 2011; Li et al., 2012; Ivanov et al., 2014) and $Co_{2.5}Mn_{0.5}TeO_6$ (CMTO; a type-I) (Mathieu et al., 2011; Singh et al., 2016). This study not only brings out structural method as a significant tool to probe low temperature magnetic anomalies, but also has evidenced the shifting of spin-lattice coupling from CTO to CMTO.

As reported previously, in case of CTO (Hudl et al., 2011 – Ivanov et al., 2012), single phase synthesis and the intrinsic low magnetic field magnetic behaviour remains under debate. The concern about this issue is its different magnetic behaviour at high and low magnetic field. In the high magnetic field, CTO shows only two antiferromagnetic (AFM) transitions at $T_{N1}$ ~ 26 K and $T_{N2}$ ~ 18 K, whereas, in the low magnetic field, it shows quite different magnetic behaviour (multiple magnetic transitions). Different groups have reported different sets of magnetic transitions including $T_{N1}$ and $T_{N2}$ (Hudl et al., 2011; Wang et al., 2013). There were no experimental reports which show all the five magnetic transitions (~ 34 K, 26 K, 21 K, 17.4 K and 16 K) in a particular ceramic or single crystal CTO grown using various techniques. Assuming the origin of these variations in its growth reaction, we have provided for the first time (Singh et al., 2016), the growth reaction mechanism of monophasic ceramic CTO followed by the experimental observation of all the five magnetic anomalies in our ceramic CTO (Singh et al., 2016). Further, we have discussed the origin of low symmetry structure of CTO in $A_3TeO_6$ (A = Mn, Ni, Cu, Co) family (Golubko et al., 2010), wherein $Mn_3TeO_6$ and $Ni_3TeO_6$ show rhombohedral symmetry, while $Cu_3TeO_6$ shows cubic symmetry. The implication of such lower symmetry structure can easily be seen in its low temperature complex magnetic structures (Hudl et al., 2011-Wang et al., 2013). Moreover, followed by the observation of mixed valence Co (high spin $Co^{2+}$ and $Co^{3+}$) in CTO which enhances the magneto-electric coupling leading to the short range magnetic interactions (Wan et al., 2016), we have reported several intriguing observations such as experimental observance of spontaneous polarization, ferromagnetic correlation below $T_{N2}$, spin phonon coupling, etc. (Singh et al., 2016). In this report, with an attempt to investigate the origin of low magnetic field (~ 5 Oe) DC magnetization behaviour and its possible correlation with the structure, we have performed low temperature synchrotron structural study. Following sections describe the magnetic behaviour at low magnetic field, their probable structural correlation followed by the discussion on the magnetic properties of CMTO.

In order to increase the spin-lattice coupling strength so as to enhance the coupling temperature in CTO, we have partially replaced Co in CTO by Mn. Mn doped CTO (CMTO) results in a single phase solid solution of CTO and MTO after a particular Mn concentration ($x \geq 0.5$) (Mathieu et al., 2011 – Singh et al., 2016). For lower Mn concentration ($x < 0.5$), two phase compounds are observed (Singh et al., 2014). As discussed above, CTO is low symmetry (C$2/c$) type-II MF material, which shows complex magnetic structure with a sequence of AFM transitions at very low temperatures. MTO, a type-I MF material, on the other hand, crystallizes in higher symmetry (R$\bar{3}$), which exhibits AFM transition at approximately the same temperature as that of CTO (Singh et al., 2016). MTO and CTO show AFM transitions at around 23 K and 26 K, respectively, as far as the high magnetic field is concerned (Ivanov et al., 2011 and Singh et al., 2016). In contrast, Mn doping in CTO enhances the AFM transition temperature to 45 K; even though AFM transition temperatures of the end members are lower (Singh et al., 2016). The preparation, characterization and the possible interpretation of enhanced anti-ferromagnetism and emergence of ferromagnetism in CMTO have been discussed previously (Singh et al., 2014 and 2016).

Literature suggests that there is no change in the crystal structures in this temperature range (5 K to 300 K) and the lattice parameters have been found to be identical within the experimental errors (Golubko et al., 2010; Ivanov et al., Toledano et al., 2012; and Wang et al., 2013). However, we feel that the change in magnetic structures at low temperatures originate most probably from a temperature dependence of the magnetic interactions and their possible correlation with the lattice. Also, we have shown in our earlier studies that mixed valence Co as well as Mn enhances the coupling strengths in CTO and CMTO (Singh et al., 2016). This indicates that one needs a precise determination of all the exchange interactions involved in CTO and CMTO, in order to achieve a deeper understanding of the low temperature magnetic properties of these compounds. We, therefore, report systematic LT-SXRD studies on CTO and CMTO, in an attempt to investigate the possible distinct changes in structural parameters at low temperature scales.

## 2. Experimental Details

Single phase polycrystalline CTO and CMTO samples have been prepared by conventional solid state reaction route in air, using $Co_3O_4$ (Alfa Aesar 99.7 %), $TeO_2$ (Alfa Aesar 99.99 %), and $Mn_3O_4$ (Alpha Easer 99.9%). Synthesis details along with their room temperature characterizations can be found elsewhere (Singh et al., 2014 and 2016). SXRD measurements have been performed on Angle Dispersive X-ray Diffraction (ADXRD) beamline (BL-12) (Sinha et al., 2013) at Indus-2 Synchrotron Light Source, India. LT-SXRD measurements are performed using transmission geometry in a liquid helium cryostat (Advanced Research Systems Inc Model No. LT3G). This set-up is capable to reach the lowest temperature of 4 K (2.7 with pumping) and the highest temperature of 450 K. The temperature is controlled using a PID temperature controller (Lakeshore 330). For the LT-SXRD measurements, samples have been cooled first to lowest temperature i.e. 4 K and then LT-SXRD data has been carried out in a warming mode. The temperature around the sample is stabilized within 0.2 K during an individual LT-SXRD measurement (with time duration of 2-3 mints). SXRD pattern is recorded on image plate based MAR 345 dtb area detector, as a 2D pattern. The two dimensional SXRD patterns acquired from area detector are integrated using the program fit2D (Hammersley et al., 1994). The refinements of the structural parameters from the diffraction patterns are obtained using Rietveld analysis employing the FULL-PROF program (Rodriguez-Carvajal, 1993). Photon energy calibrations ($\lambda$=0.82715 Å for CTO and $\lambda$=0.8269 Å for CMTO) have been done by taking SXRD pattern of $LaB_6$ NIST standard. Energy resolution ($\Delta E/E$) is estimated to be 1.5 x $10^{-4}$. DC magnetization is measured in the temperature range of 5 K - 300 K using a SQUID magnetometer (M/s. Quantum Design, model MPMS), under zero field cooled (ZFC) and field cooled (FC) conditions. Presently, ZFC magnetization has been emphasized because we have performed the structural measurements in the same mode, as discussed above. The sample has been taken in a clear gelatin capsule which contributes insignificant diamagnetic background.

## 3. Results and discussion

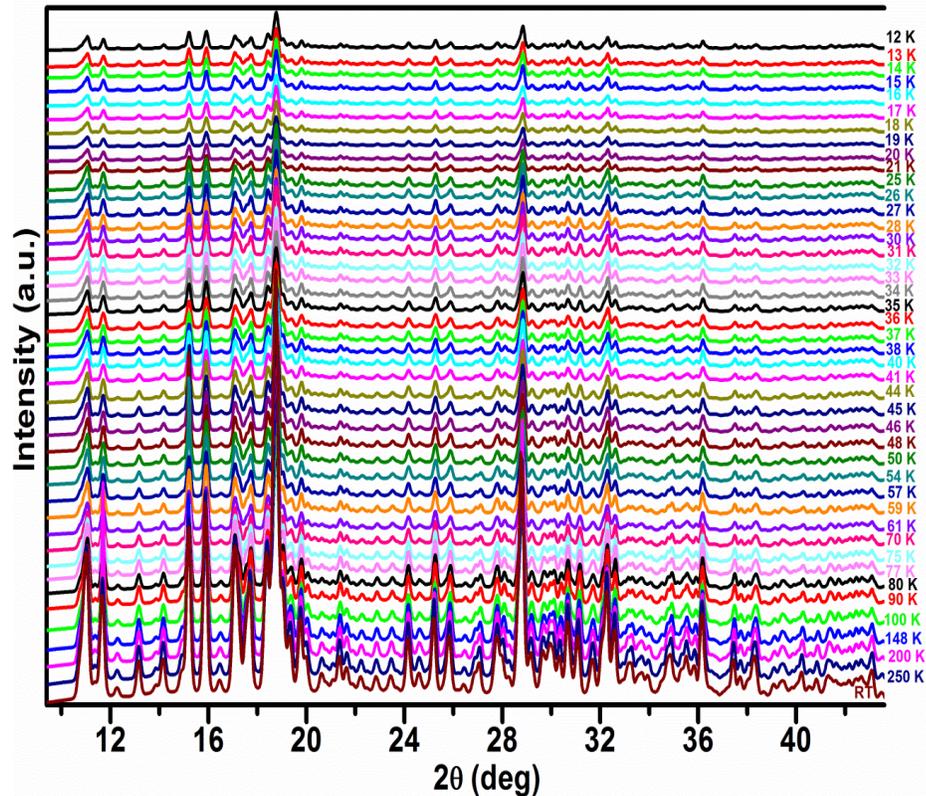

Fig.1. Systematic low temperature synchrotron X-ray diffraction patterns, which have been shifted in the vertical scale for clarity, of CTO at λ=0.82715 Å.

Figs. 1 and 2 show systematic temperature dependence of SXRD patterns of polycrystalline CTO and CMTO samples, respectively. One can see that there is no structural transition in both the sample's pattern, as far as reflections are concerned. This is in agreement with the earlier reports (Ivanov et al, 2012; Hudl et al., 2016). The only difference that one can observe is the change of intensities of the peak as well as the shift of the same in the x-axis, a common trend of materials at low temperature scale. To go into detial, we have carried out precise temperature dependent SXRD measurements i.e. with a temperature step of 1 K (Figs. 1 and 2). Also, as both CTO and CMTO show low temperature magnetic anomalies, we have focused mainly on the low temperature region. Further, in order to extract the information contained in LT-SXRD data, detailed Rietveld refinement analysis has been carried out on each data.

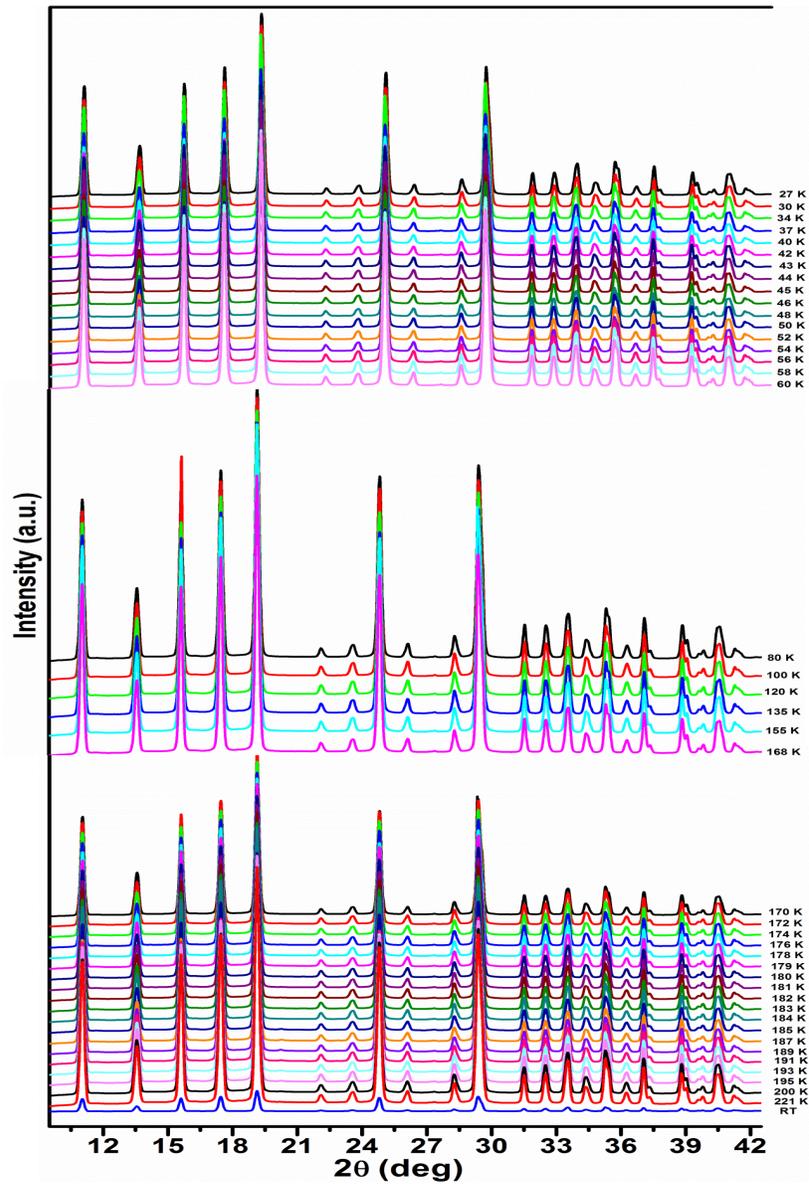

Fig. 2. Low temperature SXRD patterns which are shifted in the vertical scale for clarity of CMTO at λ=0.8269 Å.

Representative (for few temperatures) Rietveld refined patterns for polycrystalline CTO and CMTO samples are shown in Figs. 3 and 4, respectively. As reported earlier (Singh et al., 2014 and 2016), room temperature structural parameters for CTO (Monoclinic: C2/c) and CMTO (rhombohedral: R$\bar{3}$) have been found as **a = 14.8061(5)Å, b = 8.8406(3)Å, c = 10.3455(4)Å, β = 94.819(2)°** and **a = 8.6398(3)Å, c = 10.4934(2)Å**, respectively. Corresponding goodness of fit parameters have also been shown in Figs. 2 and 3. The above observation of absence of structural transition in both the samples is also validated through these Figs. 3 and 4. Further, the observed changes in the LT-SXRD patterns are quite weak (visibly). Therefore, further discussion has been drawn on the outcomes of Rietveld refinement. Fig. 5 shows lattice parameters behaviour of CTO as a function of temperature. Corresponding room temperature structural parameters can be found elsewhere (Singh et al., 2014 and 2016). In Fig. 6, we also show the corresponding magnetic anomalies observed for CTO, in order to make comment on the sensitivity of SXRD technique, against the magnetic behaviour. In order to comment about the anomalies observed in the structural parameters, we briefly mention the magnetic behaviour of CTO reported by several groups.

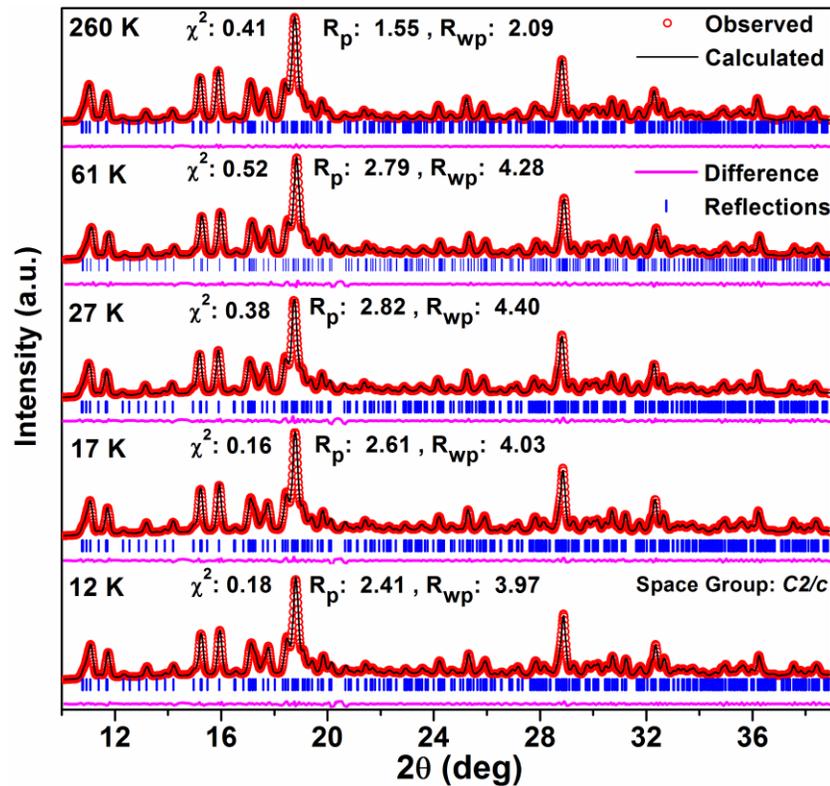

Fig. 3. Re-presentative Rietveld refinement analysis on low temperature SXRD data of CTO. Data are shifted in the vertical scale for clarity.

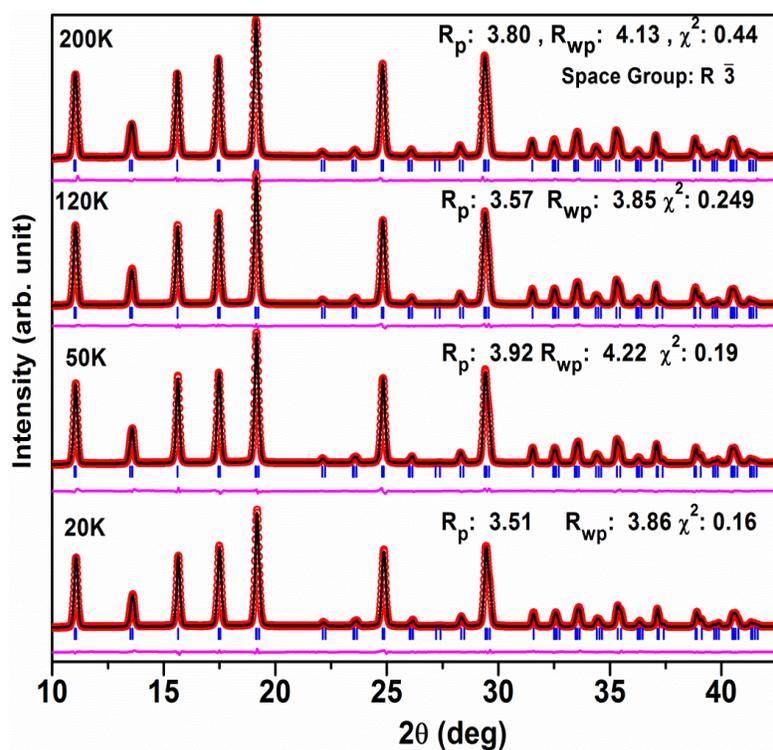

Fig. 4. Systematic Rietveld refinement analysis on few low temperature SXRD data of CMTO. Data are shifted in the vertical scale for clarity.

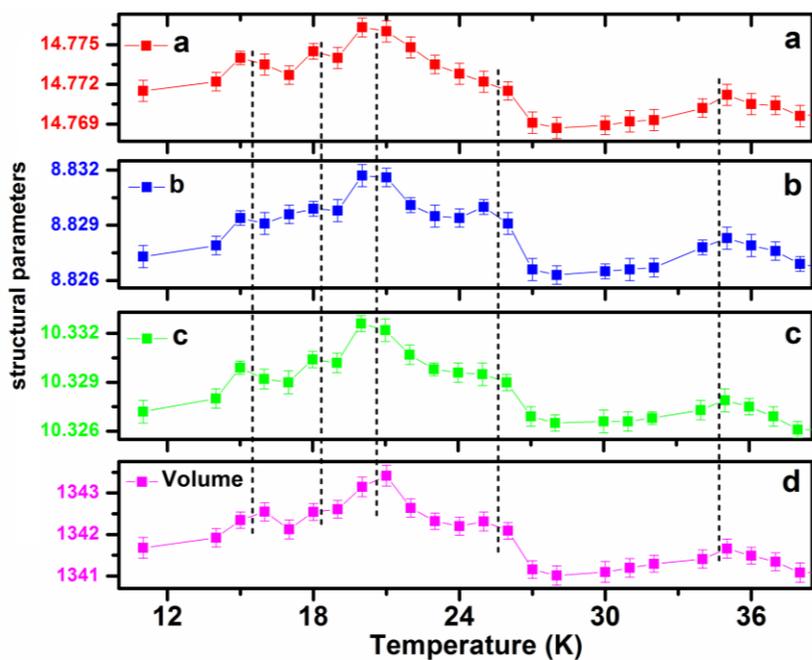

Fig. 5. Outcomes of detailed Rietveld refinement on LT-SXRD data of CTO as a function of temperature, which include all the lattice parameters (in Å) and volume (in Å$^3$), which are vertically shifted for clarity. Errors in the structural parameters have also been incorporated.

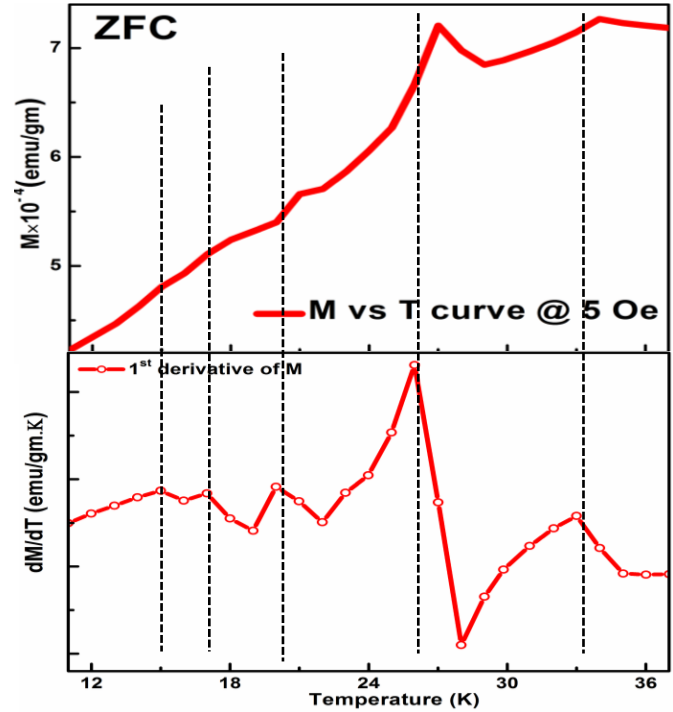

Fig. 6. Origin of the anomalies in all the structural parameters can be seen from the present magnetic behaviour of CTO. Zero field cooled magnetization (above panel) along with its first derivative plot taken at 5 Oe is shown.

The first Neutron diffraction studies (Ivanov et al., 2011) on CTO indicate first-order multi-k phase transitions; with a sequence of three AFM phases (at 17.4 K, 21 K and 26 K all below 30 K) accompanied by magnetoelectric effect. The incommensurate (IC) AFM structure emerges at $T_{N1}$ = 26 K (denoted as phase I), whereas the two commensurate AFM phases appear at 21 K (phase II) and $T_{N2}$ = 17.4 K (phase III) (Toledano et al., 2012). The evolution of phase II from phase I take place through the strong first order transition. Theoretical Landau free energy analysis, taking care of irreducible representations and monoclinic magnetic group symmetry, suggests that a strong magneto-elastic effect may be involved (Toledano et al., 2012). This would also suggests a possible change in interatomic spacing causing first order transition, which may in turn influence exchange energy from phase I to phase II. In contrast, there is a smooth second order transition from phase II to III indicating no significant discontinuity in structural parameters (Ivanov et al., 2011 and Toledano et al., 2012). Further, there are reports, which show large variations in the number of magnetic transitions in CTO. For example, Hudl et al., 2011 and Her et al., 2011 report two main magnetic transitions ~ 18 K and 26 K, in addition to 34 K and 16 K, respectively. Wang et al., 2013, on the other hand, not only report IC nature of all these magnetic transitions but also more in number (a total of four i.e. 26 K, 20 K, 18 K, and 16 K below 30 K). In summary, not only the experimental observation of all the magnetic transitions (~ 34 K, 26 K, 21 K, 17.4 K and 16 K) was necessary, but also the structural correlation to these transitions has not been reported so far. Recently, we have shown that all the reported magnetic transitions (by various groups) are present in our single phasic ceramic CTO (Singh et al., 2016). Herein, we present their structural correlation through Rietveld analysis on the temperature dependent SXRD data. Fig. 5 shows the outcomes of Rietveld refinement analysis, in terms of structural parameters (a, b, c and volume). The similar trend has also been observed for the monoclinic angle (not shown here). Further, as per our discussion about the magnetic property of CTO, Fig. 6 clearly shows all the magnetic anomalies listed above for CTO. One can see from Figs. 5 and 6 that the anomalies observed in both the structural parameters and magnetization occurs nearly at the same temperatures.

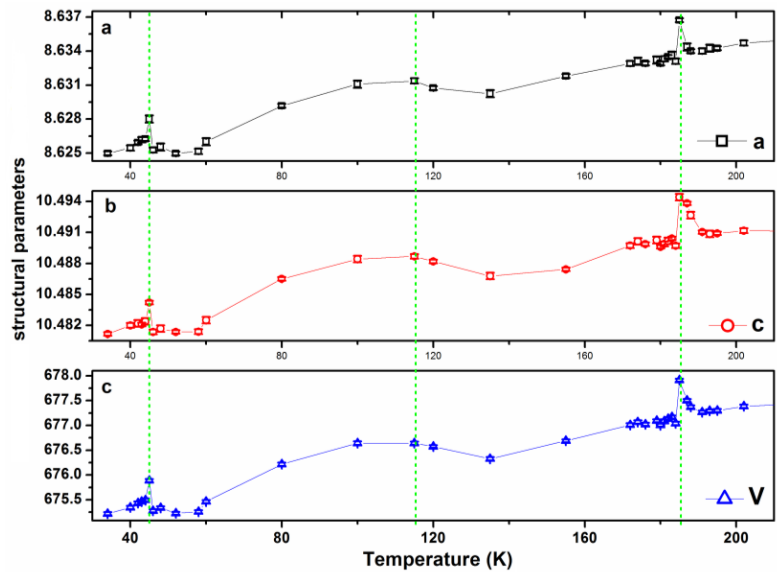

Fig. 7. Variation of lattice parameters (Å) and volume (Å$^3$) as a function of temperature, obtained from thorough Rietveld refinement on each LT-SXRD data of CMTO. Errors in the structural parameters have also been incorporated.

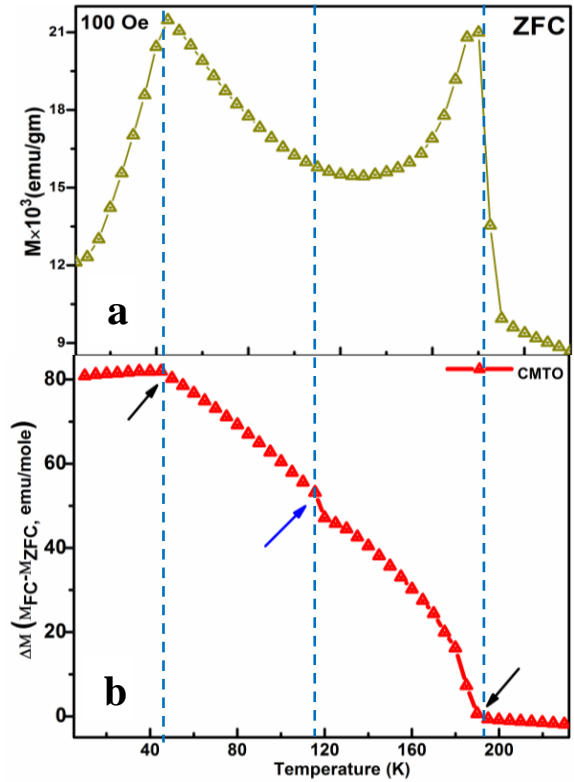

Fig. 8. Magnetic behaviour of CMTO, which contains zero field cooled magnetization (a) and its first derivative at 100 Oe (b).

As discussed above, mainly the transitions at ~ 26 K and ~ 18 K show the larger changes as compared to that observed for rest of the transition temperatures (see Fig. 5). The change in the latter one may be correlated with the change in the dielectric constant, wherein measured dielectric constant changes its trend showing a sharp upward turn forming a peak structure at around 17.4 K in zero magnetic field as the temperature is lowered below 21 K (Singh et al., 2016). The rest of the anomalies, as also observed by several groups separately, can be visualized very clearly in the temperature dependence of lattice parameters (see Fig. 5).

Similar to CTO, we also show lattice parameter variations with temperature and the observed magnetic anomalies for polycrystalline CMTO sample, respectively, in Figs. 7 and 8. The corresponding room temperature structural details can be found elsewhere (Singh et al., 2014 and 2016). In this case, we see a sharp jump in the magnetization data followed by the two kinks. The transition at 185 K and 115 K are ferromagnetic like, whereas the lower one (45 K) is an antiferromagnetic (Singh et al., 2016). It is interesting to note that the transition at 115 K, which is not visible in the ZFC magnetization, can easily be observed in the temperature variation of lattice parameters of CMTO sample, in the difference plot of FC and ZFC one. One can also sense the nature and type of magnetic interactions through the relative changes of the structural parameters variations in a material. For example, in Fig. 7, the variation at ~ 185 K is large as compared to the variation noted at ~ 45 K. In Figs. 5 and 7, we have also shown the corresponding cell volume of CTO and CMTO, respectively.

Followed by the discussions on the magnetic properties, one can see a clear signature of anomalous changes in all the structural parameters of CTO and CMTO at each magnetic transition. Depending on the nature of magnetic transitions in these two samples, variations in the lattice parameters follow the similar trend indicating the strong coupling between the spin and the lattice. Volumes also follow the similar trend as the rest of the structural parameters show. This undoubtedly shows a significant tool for probing the spin lattice couplings. Through this, one can not only investigate the structural phases exhibited by the material but can also see the effect of changes in the lattice parameters on the change in magnetic behaviour as a function of temperature. As can be seen in Fig. 6, significant anomaly at 26 K which has been attributed to the in-commensurate nature in CTO, is also visible in the present large variation in the lattice parameters of CTO (see Fig. 5).

The above discussion is based on the magnetic data observed at low magnetic fields, which has a different feature at high field as discussed in the introduction section. Following the same, CTO shows two transitions i.e. 26 K and 18 K, whereas CMTO shows only one transition at 45 K. This clearly shows large enhancement in the spin-lattice coupling. One to one correspondence in the M-T data and lattice parameters variations directly implies the sensitivity of SXRD technique to the magnetic anomalies. Further, variations in the lattice parameters with temperature for both CTO and CMTO show shifting nature of spin lattice coupling from CTO (26 K) to CMTO (45 K). All in all, a significant probing tool has been provided, which not only shows the variation in their structural parameters but also shows the possible changes in the magnetic behaviour as the materials cool down to liquid helium temperature. It is also interesting to note that the presented systematic LT-SXRD data has been carried out during the warming of the sample i.e. we first cooled the sample and then we started heating the same as in the case of ZFC magnetization measurements. This may also be the cause for the same trend of structural parameters as that of magnetization of CTO and CMTO samples.

4. Conclusions

We report systematic temperature-dependent structural and magnetic studies on CTO and CMTO, using low temperature Synchrotron X-ray diffraction and DC magnetization measurements. CTO, a low symmetry (C*2/c*) type-II multiferroic material, shows multiple magnetic transitions at low temperatures. CMTO, on the other hand, crystallizes in higher symmetry (R$\bar{3}$) type-I multiferroic material, exhibits similar but with higher AFM transition temperature along with an additional FM correlation at 185 K. Through LT-SXRD study, we have very clearly demonstrated that both the samples show strong coupling between the spin and lattice degrees of freedom. Structrural parameters obtained from LT-SXTD data show anomalies at all the magnetic transitions suggesting SXRD technique as a significant structural tool for probing spin lattice coupling.

**Acknowledgments**: Authors acknowledge Dr. P. A. Naik and Dr. P. D. Gupta for their support and encouragement. Authors sincerely thank Dr. T. V. Chandrasekhar Rao and Champalal Prajapat for providing the magnetic data on CTO and CMTO polycrystalline samples, respectively. HS thanks Dr. S. M. Gupta and Dr. Haranath Ghosh for the fruitful discussions.